# Magneto-superconductivity of "100-atm $O_2$-annealed" $RuSr_2Gd_{1.5}Ce_{0.5}Cu_2O_{10-d}$

V.P.S. Awana, and E. Takayama-Muromachi*

*Superconducting Materials Center, National Institute for Materials Science, 1-1 Namiki, Tsukuba, Ibaraki 305-0044, Japan*

M. Karppinen and H. Yamauchi

*Materials and Structures Laboratory, Tokyo Institute of Technology, Yokohama 226-8503, Japan*

Studied 100-atm $O_2$-annealed" $RuSr_2Gd_{1.5}Ce_{0.5}Cu_2O_{10-\delta}$ (Ru-1222) compound crystallized in a tetragonal *I4/mmm* space group crystal structure. Thermo-gravemetric (TG) analysis of the compound showed the release of oxygen and breaking to metallic constituents in two distinct steps at around 350 and 500 $^0$C. The DC magnetization data (*M* vs. *T*) revealed magnetic transition at ~100 K followed by superconducting transition at ~40 K. Low field ($-100 = H = 100$ Oe) *M* vs. *H* hysteresis loop showed a lower critical field ($H_{c1}$) value of around 25 Oe. Ferromagnetic component is evidenced at 5, 10, 20 and 40 K. Near saturation field of above 5 Tesla is observed at 5 K. Zero-field returning moment ($M_r$) and zero-moment coercive field ($H_c$) values at 5 K are 0.35$\mu_B$ and 250 Oe. The resistance vs. temperature (*R* vs. *T*) behaviour of the sample confirmed superconductivity at around 43 K. Superconductivity transition ($T_c$) is broadened under magnetic field with strong granularity like steps.

## I. INTRODUCTION

Coexistence of superconductivity and magnetism in ruthenium copper oxides was reported for the first time for $RuSr_2(Gd,Sm,Eu)_{1.6}Ce_{0.4}Cu_2O_{10-\delta}$ [Ru-$1^{(Sr)}2^{(Gd,Sm,Eu,Ce)}22$ or Ru-1222] [1, 2], and later for Ru-$1^{(Sr)}2^{(Gd)}12$ (Ru-1212) [3-5]. Both ruthenium copper oxides were synthesized and studied for their transport properties already in 1995 [6], but magnetic characterization had not been carried out until recently [1-5]. Both Ru-1222 and Ru-1212 phases are structurally related to the $CuA_2QCu_2O_{7-\delta}$ [Cu-$1^{(A)}2^{(Q)}12$ or Cu-1212, e.g. $CuBa_2YCu_2O_{7-\delta}$] phase with Cu in the charge reservoir replaced by Ru such that the Cu-O chain is replaced by a $RuO_2$ sheet. In the Ru-1222 structure furthermore, a three-layer fluorite-type block instead of a single oxygen-free $R$ (= rare earth element) layer is inserted between the two $CuO_2$ planes of the Cu-1212 structure [7]. In both Ru-1222 and Ru-1212 replacement of Cu in the charge-reservoir block by the higher-valent Ru increases the overall oxygen content [1-5].

Substantial work has been carried out on the Ru-1212 phase. The magnetic structure was studied through neutron diffraction experiments [8]. Electromicroscopic works revealed a super-structure along the *a-b* plane due to tilting of $RuO_6$ octahedra [9], which was further confirmed from neutron diffraction studies [10]. The appearance of bulk superconductivity at low temperatures in Ru-1212 was initially criticised [11] which was later refuted [12, 13]. Also not all single phase Ru-1212 exhibit the co-existence of superconductivity and magnetism [14]. Worth mentioning is the fact that there exist contradictory reports on heat capacity data under magnetic fields, i.e. $C_P(H,T)$ [15,16].

Even though "magneto-superconductivity" was first realised in Ru-1222, a lot still remain about the physical characterization. In particular the magnetic structure from neutron diffraction experiments and the micro-structural details are still lacking. In present contribution we report the enhancement of the superconductivity with high pressure oxygen annealing of Ru-1222, and discuss its various physical properties. TG analysis of the compound in 95% Ar and 5% $H_2$ showed the release of oxygen in two distinct steps at around 350 and 500 $^0$C.

## II. EXPERIMENTAL

The $RuSr_2Gd_{1.5}Ce_{0.5}Cu_2O_{10-\delta}$ (Ru-1222) sample was synthesized through a solid-state reaction route from $RuO_2$, $SrO_2$, $Gd_2O_3$, $CeO_2$ and CuO. Calcinations were carried out on the mixed powder at 1000, 1020, 1040 and 1060 $^0$C each for 24 hours with intermediate grindings. The pressed bar-shaped pellets were annealed in a flow of high-pressure oxygen (100 atm) at 420 $^0$C for 100 hours and subsequently cooled slowly to room temperature [17]. X-ray diffraction (XRD) patterns were obtained at room temperature (MAC Science: MXP18VAHF[22]; Cu$K_\alpha$ radiation). Thermogravimetric (TG) analysis (MAC Science: TG-DTA 2000 S) was carried out in an 95% Ar and 5% $H_2$ atmosphere to check the oxygen-stoichiometry stability. Magnetization measurements were performed on a SQUID magnetometer (Quantum Design: MPMS-5S). Resistivity measurements under applied magnetic fields of 0 to 7 T were made in the temperature range of 5 to 300 K using a four-point-probe technique.

## III. RESULTS AND DISCUSSION

### III A. X-ray diffraction and TG; Phase formation and Phase decomposition:

As-synthesized Ru-1222 copper oxide crystallises in a tetragonal structure of space group *I4/mmm* with $a = b = 3.8327(7)$ Å and $c = 28.3926(8)$ Å. An x-ray diffraction pattern for the oxide is shown in Fig. 1. A trace of $SrRuO_3$ is seen, as marked on the pattern. Presence of small amounts of $SrRuO_3$ and/or $GdSr_2RuO_6$ in Ru-1222 material is seen earlier also in various reports [1, 2, 18, 19]. In fact our currently studied sample seems to be of batter quality in terms of the presence of phase purity. Ru-1222 is structurally related to the $CuA_2QCu_2O_{7-\delta}$ [Cu-1$^{(A)}$2$^{(Q)}$12 or Cu-1212, e.g. $CuBa_2YCu_2O_{7-\delta}$] phase with Cu in the charge reservoir replaced by Ru such that the Cu-O chain is replaced by a $RuO_2$ sheet, furthermore, a three-layer fluorite-type block instead of a single oxygen-free *R* (= rare earth element) layer is inserted between the two $CuO_2$ planes of the Cu-1212 structure [7].

TG curve of the Ru-1222 compound being recorded in 95% Ar and 5% $H_2$ atmosphere with 1 $^0$C per minute heating schedule is shown in the Inset of Fig.1. Upon heating in flowing $H_2$/Ar gas Ru-1222, releases oxygen while decomposing to Ru metal, Cu metal and oxides of Sr, Gd and Ce in two distinct steps about 300 and 450 $^o$C. Owing to the sharpness of the weight loss behavior such reductive decomposition carried out in a thermobalance may be utilized in precise oxygen content determination for the compound [20]. For the Ru-1222 phase, several repeated experiments revealed that the oxygen content of presently studied 100-atm $O_2$-annealed samples was 9.63(5) per formula unit.

### III B. Electrical and Magneto-transport results:

Fig. 2 shows the resistance versus temperature plots (*R* vs. *T*) in 0, 1, 3, 5 and 7 T fields for Ru-1222. The *R* vs. *T* behaviour in zero field is metallic down to 270 K and semiconducting below 270 K until superconductivity starts with the superconductivity transition onset temperature ($T_c^{onset}$) at 47 K and the zero-resistance temperature ($T_c^{R=0}$) at 43 K. In fact the *R* vs. *T* plot above $T_c^{onset}$ is nearly temperature independent. The *R* vs. *T* behaviour under an applied field of 7 T is nearly the same as that of 0 T above $T_c^{onset}$. However in 7 T field $T_c^{R=0}$ is observed only at 12 K. In intermediate fields of 1, 3 and 5 T, $T_c^{R=0}$ were seen respectively at 18, 16 and 14 K, respectively. The $T_c^{R=0}$ value is decreased fast from 43 K to around 18 K with applied field of 1 T, and is later not affected much with higher fields of 3, 5 and 7 T. Also seen is a strong step like shoulder in the *R* vs. *T* curve at 1 T, the origin of which may be related to the intra-grain phase-lock transition. The intra-grain phase-lock transition and the appearance of bulk superconductivity below the length scale of grain size (typically 2-6μm) is discussed in detail for another magneto-superconductor $RuSr_2EuCu_2O_8$ [21]. The $T_c^{onset}$ of the compound remains nearly unchanged with *H*. For conventional HTSC (high-temperature superconductors) $T_c^{onset}$ remains nearly the same under all possible applied fields, but $T_c^{R=0}$ decreases and the transition width ($T_c^{onset} - T_c^{R=0}$) increases. In the inset of Fig. 2 the magneto-resistance (MR) data of the present Ru-1222 compound is shown at various temperatures and fields, revealing a small negative MR effect in the whole temperature range. Below 100 K the degree of MR is nearly the same in all applied fields and the nature of the MR effect is of the tunnelling-magneto-resistance (TMR) type as judged from the curve shape. Also note that the MR behaviour of the present Ru-1222 sample is different from that of Ru-1212. Ru-1212 had exhibited systematic changes in sign of MR at various $T$ and $H$ [4, 16].

### III C. Superconductivity and magnetism

Fig. 3 shows the magnetic susceptibility ($\chi$) vs. $T$ behaviour in the temperature range of 5 to 200 K for Ru-1222 sample under applied fields of 5, 10 and 50 Oe, measured in both zero-field-cooled (ZFC) and field-cooled (FC) modes. In an applied field of 5 Oe, the $\chi$ vs. $T$ show the branching of zero-field-cooled (ZFC) and field-cooled (FC) curves at around 90 K ($T_{irr}$), a step like structure in both at around 40 K ($T_c$) and further a diamagnetic transition around 40 K ($T_d$) in the ZFC magnetization. Though the ZFC and FC magnetization branching is seen at around 90 K, the magnetic behaviour starts deviating from normal paramagnetic relation at much higher $T$ say 160 K. The characteristic temperatures $T_{irr}$, $T_c$, and $T_{mag}$ are weakly dependent on $H < 100$ Oe. For higher $H > 100$ Oe, both ZFC and FC are merged with each other, and only $T_{mag}$ could be seen, see inset Fig. 3. This is in general agreement with earlier reports [1, 2, 21,22]. In fact no ZFC - FC branching is observed down to 5 K in both 1,000 and 10,000 Oe fields and both the anomaly and the irreversibility in both ZFC and FC branches look to be washed out.

The ZFC curve did not show any diamagnetic transition ($T_d$) in $H > 50$ Oe. The magnetization data at $H = 10$ Oe, show nearly the same characteristics as for $H = 5$ Oe. A low field (-100 Oe = $H$ = 100 Oe) $M$ vs. $H$ loop for currently studied Ru-1222 compound is shown in Fig.4. Interestingly the diamagnetic signal starts decreasing above applied fields of 25 Oe, and turns to zero at say 40 Oe. The compound seems to have a lower critical field ($H_{c1}$) of around 25 Oe. Interestingly the $M$ vs. $H$ plot shown in Fig. 4 does not appear to be a normal HTSC case. We will discus the low field (-100 Oe = $H$ = 100 Oe) $M$ vs. $H$ loop of Fig.4 again after further magnetic characterization in next section.

To elucidate the magnetic property of Ru-1222 we show isothermal magnetization ($M$) vs. applied field ($H$) behaviour at various $T$ (Fig. 5). Clear $M$ vs. $H$ loops are seen at 5, 10, 20, and 40 K. The applied fields are in the range of -2000 Oe = $H$ = 2000. At 5 K, the returning moment ($M_{rem}$) i.e. the value of magnetization at zero returning field and the coercive filed ($H_c$) i.e. the value of applied returning field to get zero magnetization are respectively 0.35 $\mu_B$ and 250 Oe. Worth mentioning is the fact that Gd (magnetic rare earth) in the compound orders magnetically below 2 K and Ce is known to be in tetravalent non-magnetic state hence the $M_{rem}$ and $H_c$ arising from the ferromagnetic hysteresis loops do belong to Ru only. Hysteresis loops are not seen for *M* vs. *H* plots above 80 K. For various hysteresis loops being observed from *M* vs. *H* plots below 80 K, the values of both $M_{rem}$ and $H_c$ decrease with *T*. The plots for both are shown in upper and lower insets of Fig. 5. Both $M_{rem}$ and $H_c$ of 0.35 $\mu_B$ and 250 Oe being observed for Ru-1222 are much higher than reported for other magneto-superconductor Ru-1212 [5, 12]. For Ru-1212 the hysteresis loops are reported quite narrow with $M_{rem}$ and $H_c$ of 0.085 $\mu_B$ and 10 Oe respectively. This indicates that in Ru-1222 the ferromagnetic domains are less anisotropic and more rigid.

The isothermal magnetization as a function of magnetic field at 5 K with higher applied fields; 70000 Oe = *H* = 70000 Oe is shown in Fig.6. The saturation of the isothermal moment appears to occur above say 5 T applied fields. The contribution from the ferromagnetic component starts to appear below 100 K. The presence of the ferromagnetic component is confirmed by hysteresis loops being observed at 5, 10, 20 and 40 K in the *M* vs. *H* plots, (see Fig. 5). Ru spins order magnetically above say 100 K with a ferromagnetic component within ($M_{rem}$, $H_c$ = 0.35 $\mu_B$, 250 Oe) at 5 K. As far the value of higher field (> 5 T) saturation moment is concerned, one can not without ambiguity extract the value for Ru contribution. Basically besides paramagnetic Gd contribution at 5 K, the contribution from Cu can not be ignored, which in an under-doped HTSC compound contributes an unknown paramagnetic signal to the system. For paramagnetic Gd contribution the theoretical plot at 5 K is shown in the inset of Fig. 6. After taking out the Gd contribution from Ru-1222 effective moment in Fig.6, a value of ~ 0.75 $\mu_B$ is obtained for effective near saturation moment of Ru. This value is lees than for $Ru^{5+}$ low spin state ordering. In Ru-1212 compound, based on various magnetization data the $Ru^{5+}$ state is reported with an effective saturation moment of nearly 1$\mu_B$ [3], which ironically differs with more recent magnetic analysis [23].

Superconductivity is seen in terms of diamagnetic transition at below $T_d$, and $T_c$ (*R*=0) at slightly higher temperature. It is known earlier that due to internal magnetic field, these compounds are in a spontaneous vortex phase (SVP) even in zero external field [24]. For $T_d < T < T_c$ the compound remains in mixed state. Hence though $R = 0$ is achieved at relatively higher temperatures the diamagnetic response is seen at much lower $T$ and that also in quite small applied magnetic ($H_{c1} < 25$ Oe) fields. Now we can understand the $M$ vs. $H$ loop being shown in Fig.4. As discussed in previous section clear ferromagnetic component is seen in the compound at 5 K. Hence at 5K both ferromagnetic and the superconducting hysteresis loops are present in the $M$ vs. $H$ magnetization data, and at low applied fields viz. -100 Oe = H = 100 Oe, the compound simply exhibit the superimposition of the both, which is the case in Fig.4.

## IV CONCLUSION

Good quality $RuSr_2Gd_{1.5}Ce_{0.5}Cu_2O_{10-\delta}$ (Ru-1222) is synthesized through solid state reaction route and superconductivity is achieved in the same up to 43 K with 100-atm-$O_2$ annealing. Ru-1222 releases oxygen in two distinct steps at about 300 and 450 $^o$C. The compound shows the clear magneto-superconductivity characteristics, in terms of bulk superconductivity and the presence of ferromagnetism at low $T$, below 40 K. At 5 K, the $M_{rem}$ and $H_c$ = 0.35 $\mu_B$ and 250 Oe are obtained for the ferromagnetic component in the compound. The resistance vs. temperature ($R$ vs. $T$) behaviour of the sample confirmed superconductivity at around 43 K, which is broadened under magnetic field with strong granularity like steps.

## ACKNOWLEDGEMENT

V.P.S.A. acknowledges the support of Prof. E. Takayama-Muromachi for providing him with the NIMS postdoctoral fellowship to work in NIMS.

**FIGURE CAPTIONS**

Figure 1. X-ray diffraction pattern observed for Ru-1222, inset shows the TG curve for the same.

Figure 2. *R* vs. *T* plots in 0, 1, 3, 5 and 7 T fields for Ru-1222, in inset is shown the magneto-resistance at various *T* and *H* for the same.

Figure 3. *c* vs. *T* plot for Ru-1222 in both ZFC and FC modes with *H* = 5, 10 and 50 Oe, the inset shows the same for *H* = 1,000 and 10,000 Oe.

Figure 4. *M* - *H* loop for Ru-1222 at 5 K with -100 = *H* = 100 Oe.

Figure 5. *M* - *H* loops for Ru-1222 at 5, 10, 20 and 40 K with -2000 = *H* = 2000 Oe. The upper and lower insets of the figure show $M_{rem.}$ vs. *T* and $H_c$ vs. *T* plots for the same.

Figure 6. *M* - *H* plot for Ru-1222 compound at *T* = 5 K, the applied field are in the range of -70000 Oe = H = 70000 Oe. The inset shows the theoretical plot for paramagnetic Gd contribution to the system.

## REFRENCES

Figure 1, Awana et al.

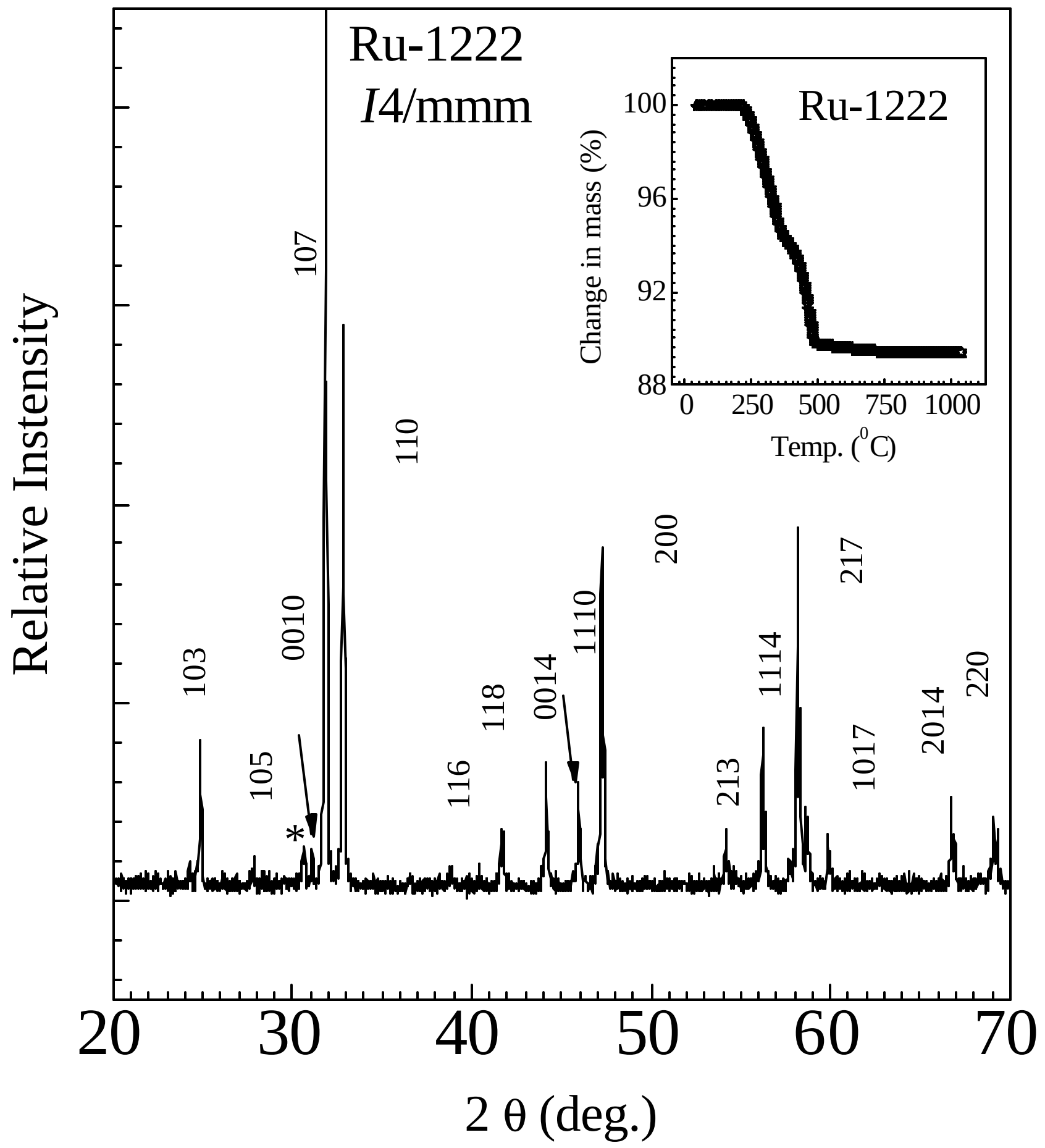

Figure 2, Awana et al.

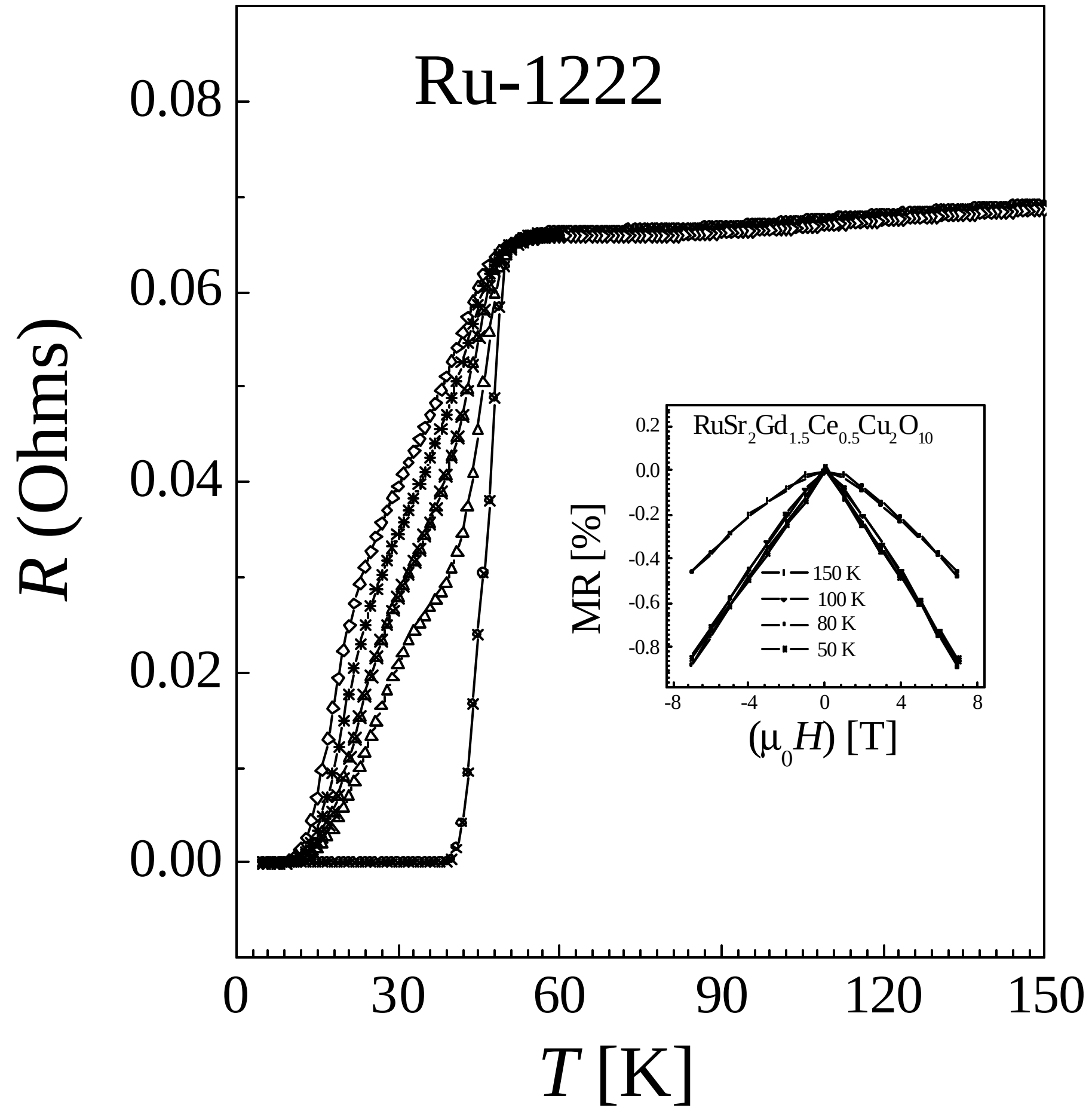

Figure 3,  Awana et al.

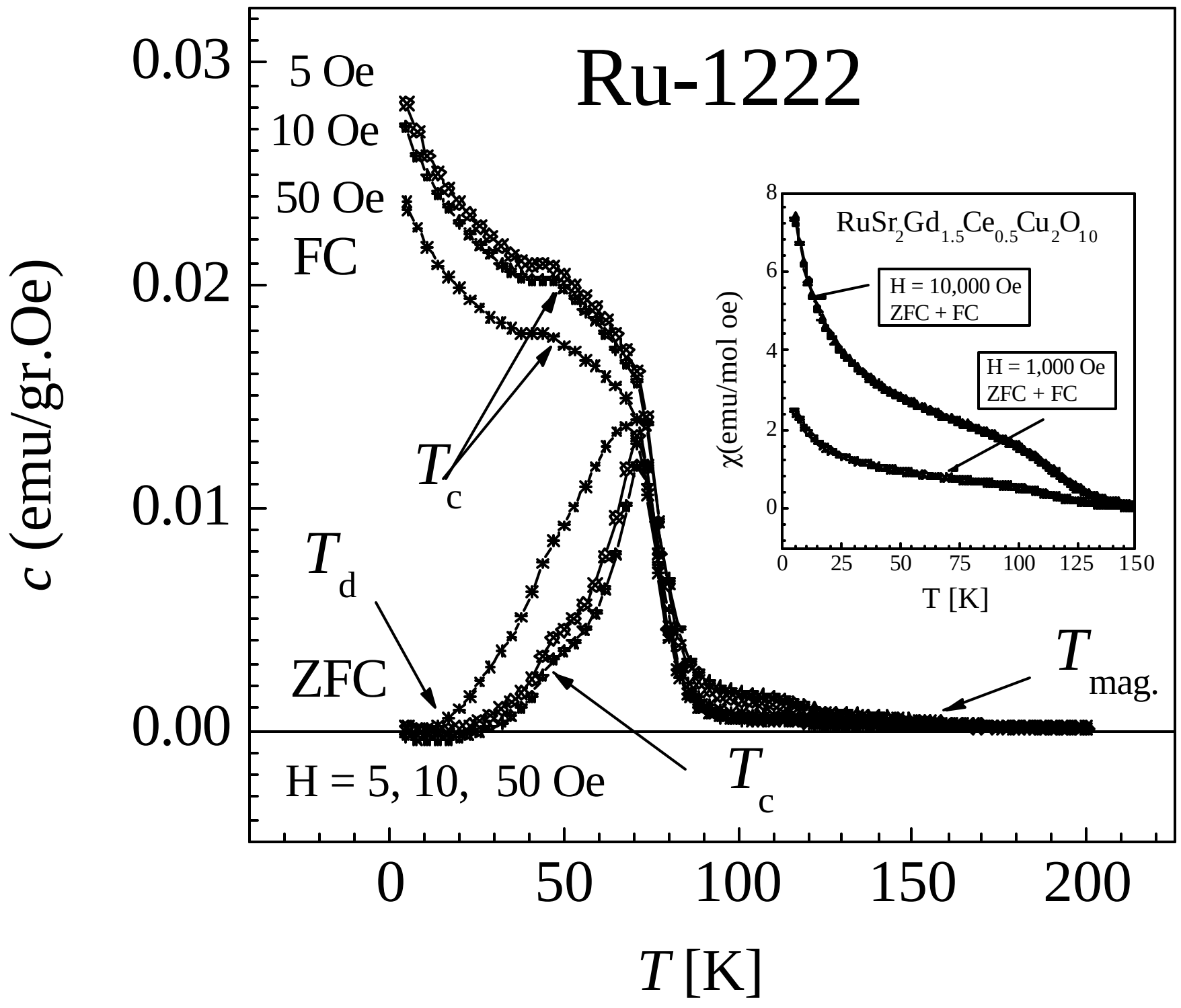

Figure 4, Awana et al.

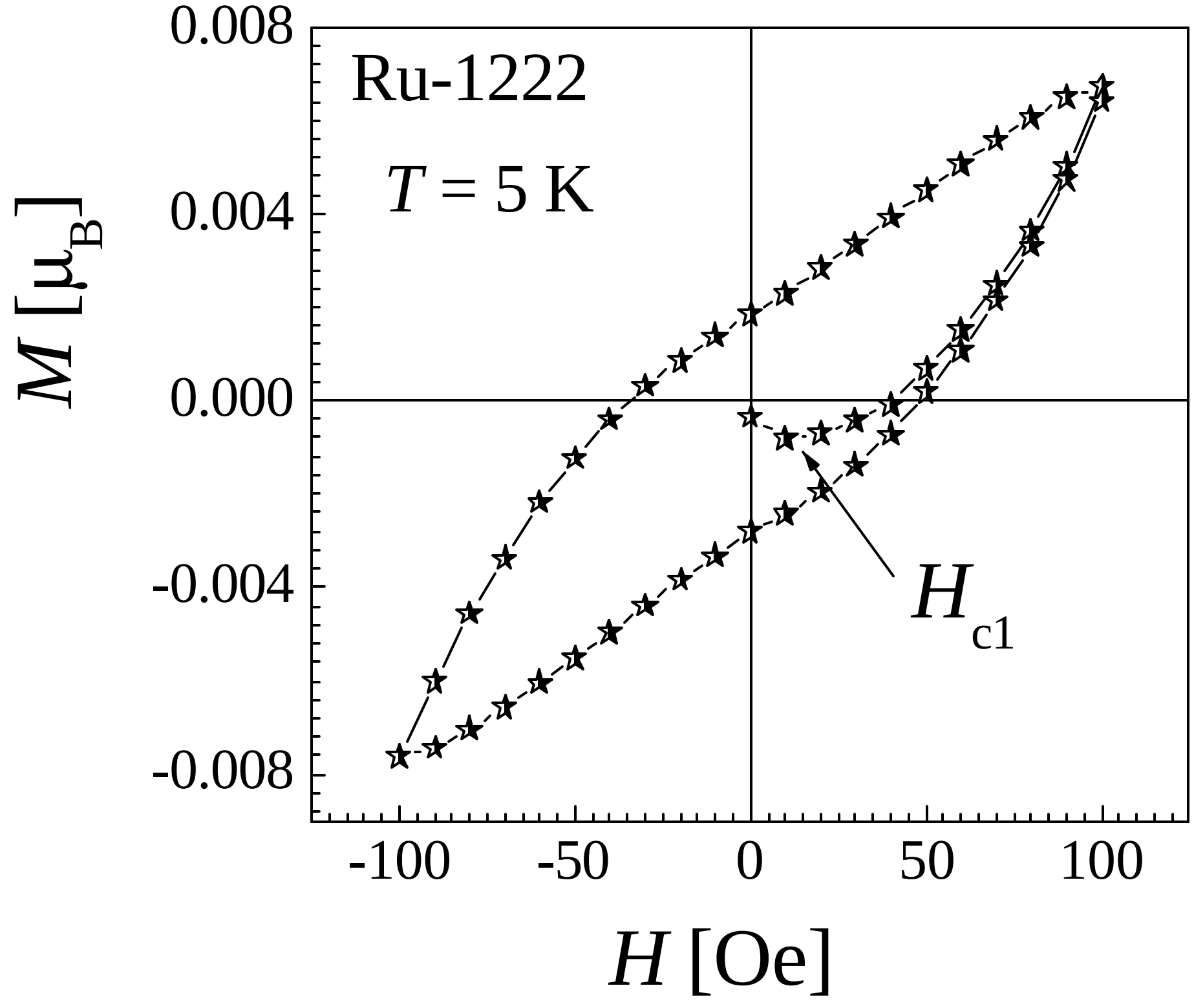

Figure 5, Awana et al

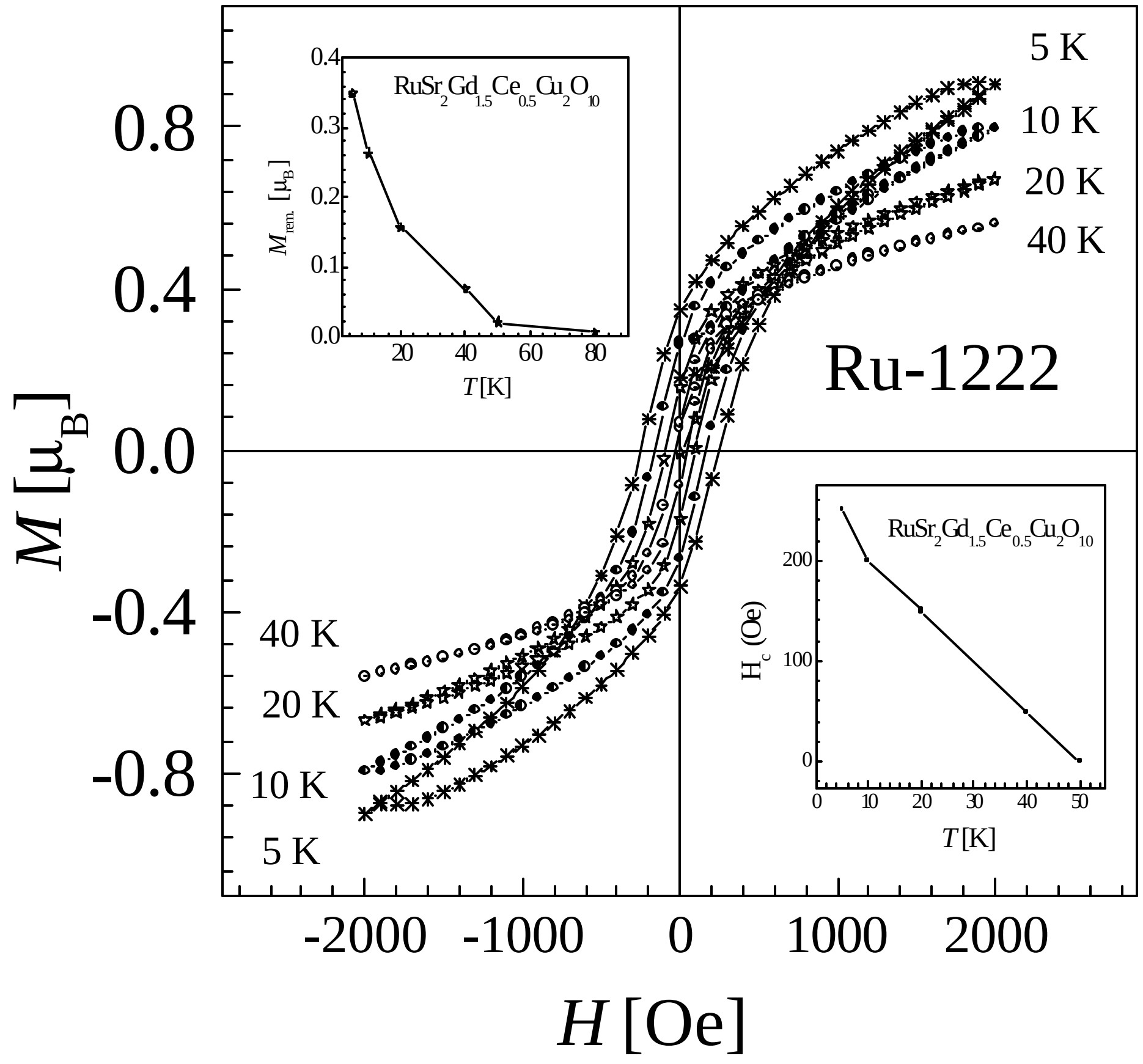

Figure 6, Awana et al

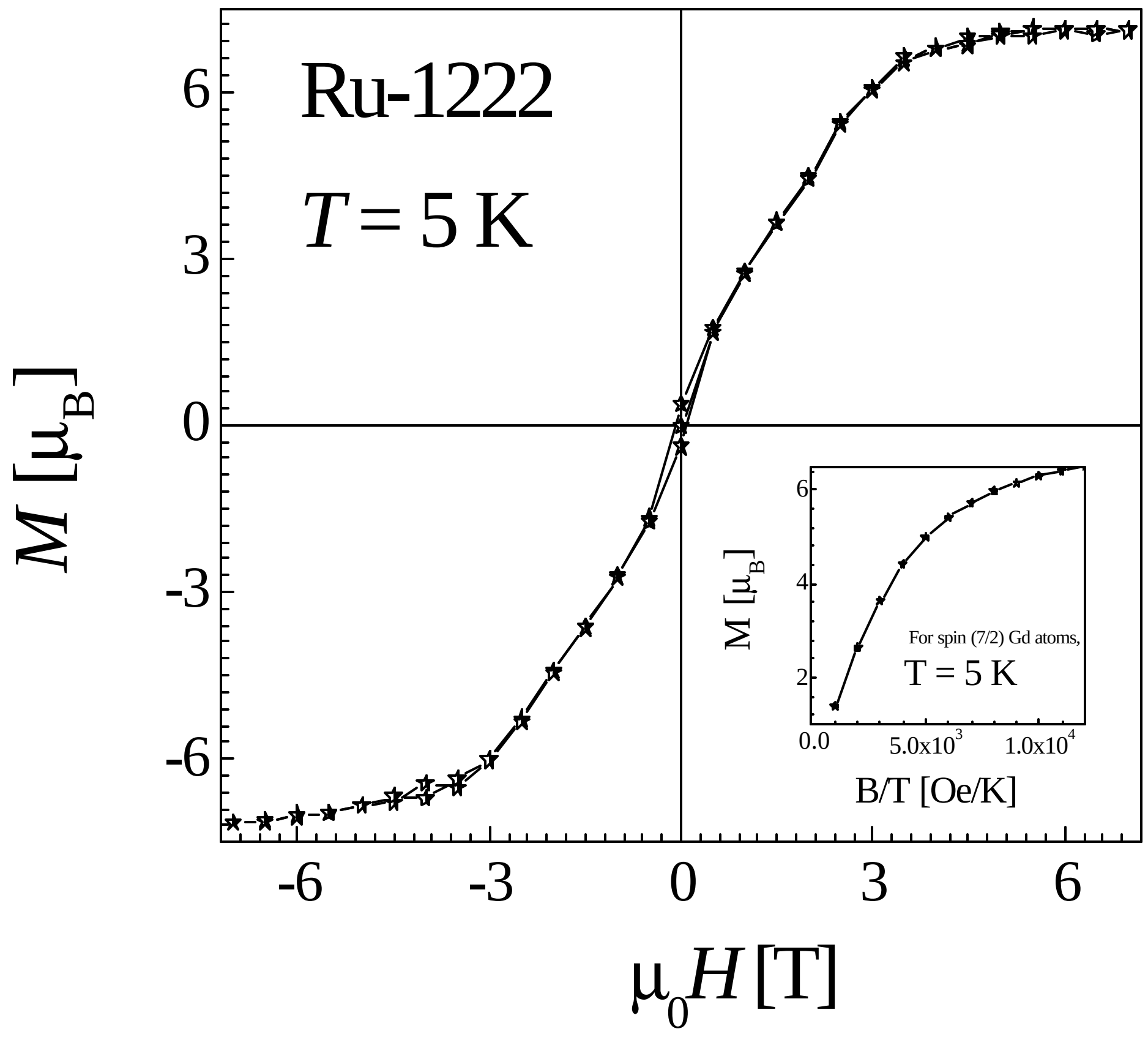